\documentclass[preprintnumbers,showpacs,twocolumn,amsmath,amssymb]{revtex4}

\usepackage{amsmath}
\usepackage{amsfonts}
\usepackage{amssymb}
\usepackage{verbatim}
\usepackage{graphicx}
\usepackage{color}

\newcommand{\beq}{\begin{equation}}
\newcommand{\eeq}{\end{equation}}
\newcommand{\bqa}{\begin{eqnarray}}
\newcommand{\eqa}{\end{eqnarray}}

\newcommand{\erf}[1]{eq.~(\ref{#1})}

\newcommand{\bra}[1]{ \langle{#1} |}
\newcommand{\ket}[1]{ |{#1} \rangle}

\newcommand{\ro}[1]{\left( {#1} \right)}

\newcommand{\s}[1]{\hat\sigma_{#1}}

\newcommand{\Str}{\mathcal{S}}

\newcommand{\xfrac}[2]{{#1}/{#2}}

\newcommand{\blk}{\color{black}}

\definecolor{ngreen}{rgb}{0.2,0.6,0.2}

\definecolor{golden}{rgb}{0.8,0.6,0.1}

\definecolor{purp}{rgb}{0.8,0.1,0.8}

\definecolor{orange}{rgb}{0.9,0.3,0}
\definecolor{mar}{rgb}{0.6,0.1,0.1}

\begin{document}

\title{Arbitrarily loss-tolerant Einstein-Podolsky-Rosen steering allowing a demonstration over 1~km of optical fiber with no detection loophole}

\author{A.\ J.\ Bennet,$^{1,2}$ D.\ A.\ Evans,$^{1,2}$ D.\ J.\ Saunders,$^{1,2}$ C.\ Branciard,$^3$ E.\ G.\ Cavalcanti,$^{2,4}$ H.\ M.\ Wiseman,$^{1,2~\dag}$ G.\ J.\ Pryde$^{1,2~ \ddagger} $\\}

\affiliation{$^{1}$Centre for Quantum Computation and Communication Technology (Australian Research Council), Griffith University, Brisbane, 4111, Australia\\ $^{2}$Centre for Quantum Dynamics, Griffith University, Brisbane, 4111, Australia\\ $^{3}$School of Mathematics and Physics, University of Queensland, Brisbane, 4072, Australia\\ $^{4}$School of Physics, University of Sydney, NSW 2006, Australia}

\date{\today}

\begin{abstract} 
Demonstrating nonclassical effects over longer and longer distances is essential  for both \blk quantum technology and fundamental science. 
The  main challenge \blk is loss of photons during propagation, because considering only those cases where photons are detected opens a ``detection loophole'' in security whenever  parties or devices are untrusted. Einstein-Podolsky-Rosen (EPR) steering is equivalent to an entanglement-verification task in which one party (device) is untrusted. We derive arbitrarily loss-tolerant tests,  enabling us to perform \blk a detection-loophole-free demonstration of EPR-steering with parties separated by a coiled 1~km optical fiber,   with a total loss  \blk of 8.9~dB ($87\%$).
\end{abstract}

\pacs{03.65Ud, 42.50Xa}

\maketitle

\section{Introduction}

In quantum mechanics, when two particles are in a  pure  entangled state, a measurement of one (say, Alice's) induces an apparent nonlocal collapse of the state of the other (Bob's),   as first discussed by Einstein, Podolsky and Rosen (EPR)~\cite{EinEtalPR35}. Schr\"odinger realized that with a maximally entangled state, for any given observable Bob chooses to measure, Alice can, by an appropriate choice of her own measurement, ``steer'' Bob's state into an eigenstate of his observable and thus predict its outcome~\cite{SchPCP35}. The recent formalization~\cite{Wiseman2007} of  ``EPR-steering''~\cite{CavJonWisRei09}  as a quantum information task  further generalizes Schr\"odinger's notion by allowing for mixed  states and imperfect measurements.

In the  EPR-steering task, Alice tries to convince Bob, who does not trust her,  that they share pairs of entangled quantum particles~\cite{Saunders:2010}. The  protocol  requires   Alice and Bob to compare results from rounds of local ``measurements'' on each pair of particles.   Bob's  measurement is always genuine, but he cannot assume that Alice's is---a dishonest Alice may instead try to cheat. The only way  for an honest Alice to distinguish herself is by  demonstrating her ability to steer Bob's state.  A dishonest Alice may employ powerful cheating strategies which to Bob would appear indistinguishable from loss, opening the ``detection loophole''. For this reason Bob cannot simply ignore cases when Alice does not (or claims not to) detect a photon. Thus there is a great challenge in verifying entanglement sharing with an untrusted party over a long distance. Using high-efficiency detectors can only ever compensate for moderate transmission losses. For high losses, what is required is a more sophisticated theoretical and experimental approach. 

In this paper, we demonstrate theoretically and experimentally that EPR-steering can be rigorously performed even in the presence of arbitrarily high losses. Other photonic protocols have been implemented in parallel with this work~\cite{Wittmann2011,Smith2011} using high efficiency sources and detectors; however, they are not arbitrarily loss tolerant --- both use at most 3 measurement settings and hence are limited to losses less than $67\%$.  (We note that Ref.~\cite{Wittmann2011} also closes the locality and freedom-of-choice loopholes~\cite{Scheidl2010}, which is of interest in fundamental tests of quantum mechanics.) Our experiment 
uses up to 16 settings, in conjunction with completely new, maximally loss-tolerant, tests, allowing 
the first   demonstration of Einstein's ``spooky action''~\cite{EinEtalPR35,Born49} over a long (1~km),  lossy (87\% loss) channel. As such, it opens the door to using detection-loophole-free EPR-steering inequalities as tools in quantum information science, such as guaranteeing secure one-way entanglement sharing.

\subsection{EPR-Steering and the Detection Loophole}

\begin{figure}
\begin{center}
\includegraphics[width=.9\linewidth]{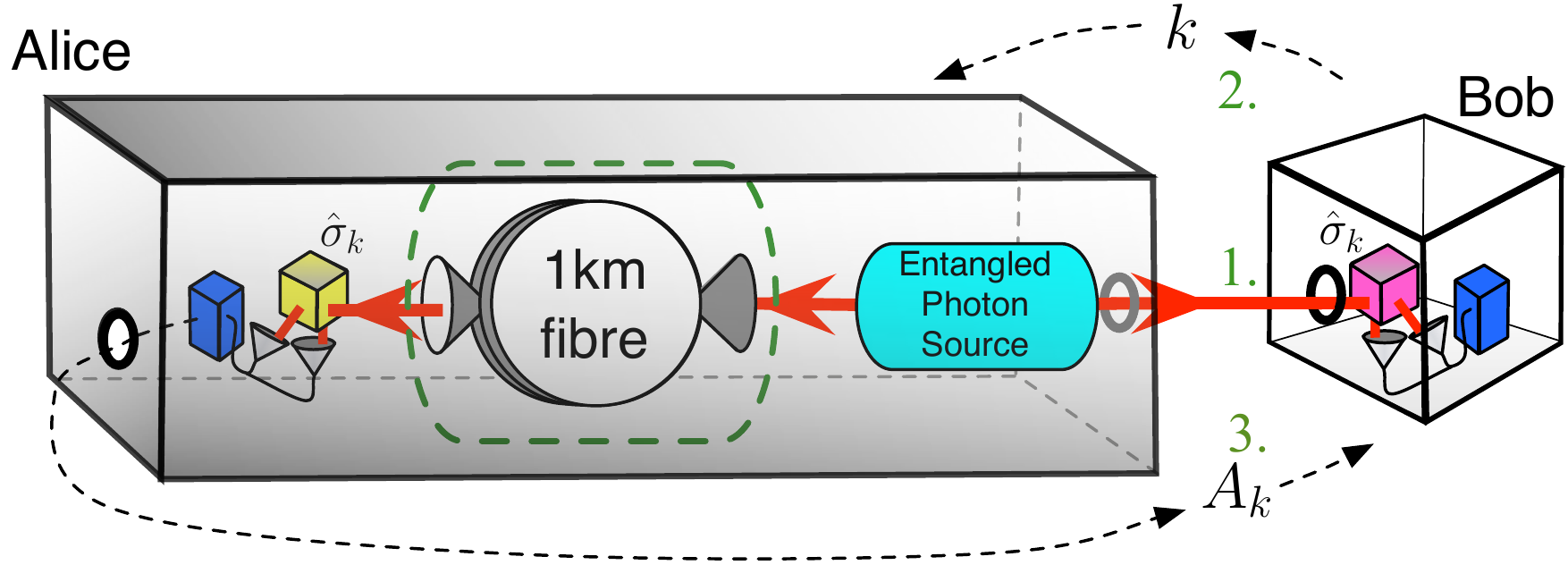}
\end{center}
\caption{ {\bf Conceptual representation of the EPR-steering task.} In each round of the protocol, {\bf 1.}~Bob receives a photonic qubit, {\bf 2.}~announces a measurement setting, $k$, and {\bf 3.}~receives a ``measurement'' result from Alice---see text for details. Bob must assume that Alice controls the source, her line, and her detectors (all enclosed in the grey box). Bob implements the measurement $\hat{\sigma}_k$ (pink cube) and monitors the measurement outcome (blue cube). In the case of an honest Alice, Bob's qubit is half of an entangled pair and Alice's measurement results are genuine; Alice measures in the same direction as Bob (yellow cube) using an identical apparatus. We demonstrate EPR-steering over 1~km of optical fiber inserted in the line on Alice's side (green dashed box).}
\label{fig:protocol}
\end{figure}

The formal procedure Bob implements to be certain he has observed EPR-steering is as follows (Fig.~\ref{fig:protocol}): {\bf 1.}~Bob receives his  quantum system {\bf 2.}~Bob announces to Alice his choice of measurement setting (labelled $k$)  from a predetermined   set  of $n$  observables. {\bf 3.}~Bob records his measurement outcome and Alice's declared result,  $A_k$. {\bf 4.}~Steps {\bf 1}--{\bf 3} are reiterated to obtain the average correlation  between Alice's and Bob's results, known as the \emph{steering parameter} $\Str_n$. If $\Str_{n}$ is larger than  a certain {\em EPR-steering bound} $C_{n}$, i.e. if it {\em violates} the {\em EPR-steering inequality $\Str_{n} \leq C_{n}$}, Alice has  successfully demonstrated EPR-steering. Such a demonstration rules out all ``local hidden state models''~\cite{Wiseman2007,CavJonWisRei09} for the observed correlations, i.e. the class of local realistic models in which Bob's system is described by a local quantum state and Alice's system by a local hidden variable.

The obvious strategy for Alice to cheat is to send Bob, in each round, a single qubit in an eigenstate of one (chosen at random) of the $n$ observables, and then decline to announce a result whenever her observable does not correspond to Bob's announced measurement. In this way, she can mimic the perfect  correlations of a  maximally entangled state on  the trials where she announces a result. Bob cannot be sure whether her unannounced results are due to cheating or to genuine loss of her qubit, e.g. by photon absorption or scattering  during transmission. He can only infer entanglement from the correlations if he makes a \textit{fair sampling assumption}, that Alice's loss events were independent of  his setting. This assumption cannot be made if Alice is untrusted, opening up a so-called ``detection loophole''~\cite{Pearle1970}. Consequently, even if  an untrusted Alice {\em is}  honest, the protocol just described cannot be used by Bob to verify entanglement with her, or to  rigorously test Einstein's ``spooky action'', over a long distance channel where losses are large. 

We use the term ``detection loophole''  for  EPR-steering because it is analogous to that for Bell inequalities~\cite{BelPHY64}. 
The latter are similar to EPR-steering inequalities except that {\em neither}  Alice {\em nor} Bob are trusted~\cite{Wiseman2007}. 
Hence the Bell detection loophole applies to both parties, while the EPR-steering detection loophole applies only to Alice. Note that the detection apparatus is part of a ``party'';  unless the apparatus is trusted, the party cannot be. Bell inequality violations have been demonstrated experimentally~\cite{Aspect1982,Tittel1998,Weihs1998,Rowe2001},  albeit with the fair-sampling assumption, or other extra assumptions~\cite{Pearle1970,Grangier2001, Branciard2011,Scheidl2010,Har98}. Violating an EPR-steering inequality is easier than violating a Bell inequality, but harder than witnessing entanglement   (with trusted parties)~\cite{Wiseman2007}.  This hierarchy has previously been demonstrated experimentally  (using the fair sampling assumption) both in terms of noise tolerance~\cite{Saunders:2010} and in terms of experimental parsimony~\cite{Saunders2011}. The same hierarchy exists in terms of how loss-tolerant these tests can be without the fair sampling assumption, and indeed the EPR-steering tests we perform here  can be made
arbitrarily loss-tolerant.

EPR-steering has previously been demonstrated in optical systems without the fair sampling assumption, using high-efficiency homodyne detectors~\cite{OuPereira1992,BowenSchnabel2003}. However, unlike the protocols introduced here, those protocols using two-mode squeezed states and quadrature measurements, in which Alice gets a result every time, cannot be used for losses greater than 50\%. The reason is that if the channel losses are greater than $50\%$, then an untrusted Alice could, as far as Bob knows, actually have a zero loss channel, and be using a 50:50 beam-splitter to effect a simultaneous measurement of both quadratures. Such a dual measurement would allow Alice to choose, after the fact, which quadrature to report as having been measured, with no actual measurement choice on her part. But measurement choice by Alice is essential to any demonstration of EPR-steering. In fact this limit of $50\%$ loss holds no matter how many different quadratures Alice may measure \cite{Jones2011,WisGam12}.

In contrast with the above continuous variable versions of the EPR paradox, and with the photonic protocols used in other recent experiments \cite{Wittmann2011,Smith2011}, the type of photonic protocol we introduce works for arbitrarily high losses. 
As a consequence, ours is the first that can overcome long-range transmission losses due to scattering in the atmosphere or, as in our experiment, absorption in optical fiber.  Indeed, our demonstration through a 1km fiber is a key technical advance, 
showing the way forward to long-range application of EPR-steering, whether for fundamental investigations of quantum mechanics or quantum communications applications.

\subsection{EPR-Steering and Quantum Cryptography}

Nonclassical effects such as Bell nonlocality and EPR-steering not only illuminate fundamental issues in quantum mechanics; they also have direct applications in quantum technology.  For instance, the security of quantum key distribution (QKD) systems requires the existence of a channel that can transmit entanglement~\cite{Curty2004}. The violation of a Bell inequality proves the existence of such a channel with no need for any assumptions about the devices involved. This allows for \textit{device-independent} (DI) secure QKD~\cite{Acin2007}: the two parties can {establish} a secret key even if they bought their equipment from an adversary. Bob's ability to verify entanglement via EPR-steering provides a similar resource for quantum communication.

Specifically, it  has recently been shown~\cite{BCSW11} that  performing an EPR-steering task allows for one-sided DI secure QKD, appropriate when Bob (at a base station, say) can trust his detection apparatus, but cannot trust that of Alice (a roaming agent). When Bob trusts his device, this provides an equivalent degree of security as that of a Bell-inequality violation~\cite{BelPHY64}. In both cases, it is essential for the security of the protocol that there be {\it no detection loophole}. By contrast, the locality and freedom-of-choice loopholes are not important in the cryptography context,  because it is a necessary assumption of security proofs  that no information escapes from Alice's or Bob's lab unless they allow it.  

\section{Loss-tolerant EPR-Steering Inequalities}

\subsection{New EPR-steering bounds as a function of Alice's heralding efficiency}

In EPR-steering, Bob trusts his own apparatus, so he can discard those experimental runs where he fails to detect a photon, without having to invoke the fair sampling assumption. Because Bob's detector settings are known only to him prior to the detection of his photon, Alice cannot exploit the  loss of Bob's photon---either inside or outside his lab. However, Bob cannot trust any claims Alice makes about the propagation losses or the efficiency of her detectors.
In particular, he does not trust Alice's claims about how often she sees a photon, conditional on his detecting one. Rather, Bob makes use only of Alice's {\it heralding efficiency} $\epsilon$: the probability that she {\it heralds} Bob's result by {\em declaring} a non-null prediction $A_k$ for it. This is a quantity determined by Bob wholly from the experimental frequencies of events to which he has direct access.

The key result of this paper is that Bob can close the EPR-steering detection loophole, even with arbitrarily high loss, by making two modifications to the EPR-steering task described above. First, he must calculate Alice's heralding efficiency $\epsilon$ from the protocol data.  Second, he must  compute a new, $\epsilon$-dependent  bound $C_n(\epsilon)$ which the steering parameter $\Str_n$ must exceed to demonstrate EPR-steering. This procedure, described in detail below, involves determining Alice's optimal ``cheating strategy'' for a given $\epsilon$. Her optimal strategy comprises probabilistic combinations of deterministic strategies. Of course if Alice ``cheats'' like this she will not actually fool Bob, as she will not violate the EPR-steering inequalities we derive. 
 
For every set of $n$ observables measured by Bob there will be a different EPR-steering inequality. Intuitively, the most useful inequalities will result from measurements that are as mutually distinct as possible. For qubits, this suggests using measurement axes regularly spaced on the Bloch sphere. Only the vertex-to-vertex axes of the Platonic solids meet this criteria, and, as in Ref.~\cite{Saunders:2010}, these will be used to define our measurement sets. The exception is $n=16$,  for which we create a geodesic solid by combining the axes of the dodecahedron ($n=10$) and its dual, the icosahedron ($n=6$)~\cite{note1}.

Bob's measurements are described using quantum observables---in this case Pauli matrices $\hat{\sigma}_k^B $ for $k\in \{1,...,n\}$---but we make no assumption about  what Alice is doing and thus represent her declared results by a random variable $A_k \in \{-1,1\}$. Generalizing the inequalities derived in Ref.~\cite{Saunders:2010}, we derive bounds $C_n(\epsilon)$  such that when the experimental statistics, post-selected on Alice's conclusive results, violate the inequality
\begin{equation}
\Str_{n} \ \equiv \ \frac{1}{n} \, \sum_{k=1}^{n} \, \langle A_k \hat{\sigma}_{k}^{B}\rangle \ \leq \ C_n(\epsilon),
\label{eq:ineq_epsilon}
\end{equation}
this demonstrates EPR-steering {\em with no detection loophole} (without relying on a fair-sampling assumption for Alice).

\subsection{Determining the EPR-steering bounds $C_n(\epsilon)$}

\label{sec_calc_bnd}

In the idealised scenario where Alice declares a non-null result for all emitted pairs of systems ($\epsilon{=}1$), the EPR-steering bound $C_n = C_n({\epsilon{=}1})$ in Eq.~(\ref{eq:ineq_epsilon}) is given by~\cite{Saunders:2010}  
\begin{equation}
C_{n}=\underset{\{A_{k}\}}{{\rm max}}\left\{\lambda_{{\rm max}}\left(\frac{1}{n}\sum_{k}A_{k}\hat{\sigma}_{k}^{B}\right)\right\},
\label{eq:ideal_bound}
\end{equation}
where $\lambda_{\rm max}(\hat{O})$ is the maximum eigenvalue of $\hat{O}$: $C_n$ is derived by considering the maximum achievable correlation when Alice sends a known (to her) state $\ket{\xi}$ to Bob. Moreover, the eigenvectors associated with the $\lambda_{\rm max}$ for every set $\{A_k\}$ that attains the maximum define the set of optimal states $\{\ket{\xi_i}\}$ which Alice can send to Bob in order to attain the bound. These are known as Alice's {\it optimal} ``cheating ensemble'' (though of course a dishonest Alice cannot actually cheat Bob). For two qubits, the maximum value of $\Str_n$ that can be achieved is unity, and this requires a maximally entangled state, while $C_n < 1$ for $n>1$ as long as Bob's settings correspond to different observables.

If, on the other hand, Alice does not always declare a result $A_k \in \{-1,1\}$ when requested, then she could be using her knowledge of Bob's state to post-select on her outcomes in a way that allows her to violate the bound $C_n$ even without entanglement; hence the need to calculate a (higher) bound $C_n(\epsilon)$.
The bounds $C_n(\epsilon)$ are, by definition, the highest correlations which Alice can achieve by using cheating strategies, in which she sends a known (to her) pure state $\ket{\xi}$ to Bob drawn from some ensemble which depends now upon both $n$ and $\epsilon$.

\begin{figure}
\begin{center}
\includegraphics[width=0.45\linewidth]{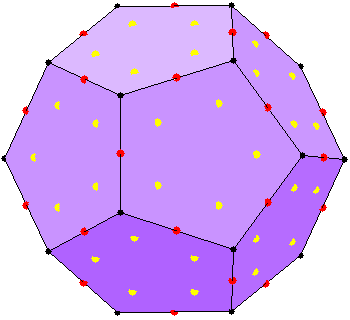} \hspace{0.05\linewidth} \includegraphics[width=0.45\linewidth]{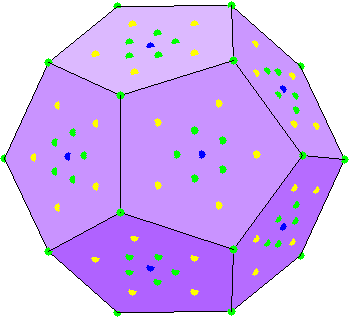}
\end{center}
\vspace{-1ex}
\caption{ {\bf Alice's optimal cheating ensembles for $n=10$ and $\epsilon \in [0.2,0.5]$.} This figure shows,  in Bloch space, the directions of the states in Alice's optimal ``cheating ensembles'' that set the bound  $C_n(\epsilon)$ for EPR-steering when $\epsilon \in [0.2,0.3]$ (on the left) and when $\epsilon \in [0.3,0.5]$ (on the right),   for the exemplary case of $n=10$. The black dots (only visible on the left) define Bob's measurement axes: the vertices of the dodecahedron. For the deterministic strategy where Alice gives non-null results for only two of Bob's settings ($m = 2$), the   red   dots (on the left) define the optimal states Alice should send,   and likewise for $m = 3$ (yellow; both), $m = 4$ (green; right) and $m = 5$ (blue; right).  For any heralding efficiency $0.2 < \epsilon < 0.3$ (left) Alice's optimal strategy is nondeterministic: a mixture of the $\epsilon=0.2$ strategy (red)  and the $\epsilon=0.3$ strategy (yellow).  For any $0.3 < \epsilon < 0.5$ (right) a dishonest Alice should use a mixture of the $\epsilon = 0.3$ (yellow) and $\epsilon = 0.5$ (blue) strategies. Interestingly, the $m = 4$ (green) deterministic strategy is {\it never} used; this is seen also in Fig.~\ref{fig:Data} (where the corresponding point does not lie on the curve representing the optimal strategy).}
\label{fig:solid}
\end{figure}

To determine Alice's optimal cheating strategies, we first consider deterministic strategies, in which Alice's declaration of +1, -1, or null as a result is determined by the state she sends, and the setting $k$ Bob specifies. For each state $\ket{\xi}$ sent to Bob, we give Alice the power to declare her outcome only if Bob requests a result for $k$ within a particular subset (depending on $\xi$)  containing $m$ elements, and declare a null result for the remaining $n-m$.  This gives an apparent efficiency $\epsilon = m/n$.  For a given $m$, the set of states which allow Alice to maximize her correlation defines a ``cheating ensemble''.
When Alice chooses states from a single cheating ensemble, she is employing a deterministic cheating strategy. In such a strategy, the optimal bound on $\Str_n$ that she can attain is given by
  \begin{equation}
D_{n}(m)=\underset{\{A_{k}\}_m}{{\rm max}}\left\{\lambda_{{\rm max}}\left(\frac{1}{m}\sum_{k}A_{k}\hat{\sigma}_{k}^{B}\right)\right\},
\label{eq:deterministic_bound}
\end{equation}
where the maximisation is over sets $\{A_k\}_m$ where exactly $m$ of the $A_k$ take values $\pm 1$, while the rest are null (and can be taken to have value $0$, for mathematical convenience).  Moreover, performing the maximization reveals the optimal cheating ensemble, as each set $\{A_k\}_m$ for which the maximum $\lambda_{\rm max}$ in \erf{eq:deterministic_bound} is attained defines a state: the eigenstate corresponding to that $\lambda_{\rm max}$. In general there are several such sets $\{A_k\}_m$ which attain the maximum in \erf{eq:deterministic_bound}, and we will use $p(m)$ to denote
their number, which is thus also the number of states in the optimal cheating ensemble for a given $m$. Examples of such cheating ensembles $\{\ket{\xi^{(m)}_i}\}_{p(m)}$ are shown in Fig.~\ref{fig:solid}, for  $n=10$ and $m \in \{2, 3, 4, 5\}$.

However, Alice does not necessarily have to choose strategies where exactly $m$ out of her $n$ measurements are non-null.
Indeed, the optimal deterministic strategies just considered are not necessarily the optimal strategies for Alice even for an apparent efficiency such that $\epsilon n$ is an integer $m$, and clearly do not apply if $\epsilon n$ is not an integer. For any $\epsilon$, we must consider Alice's most general strategy: a probabilistic mixture of optimal deterministic strategies of different $m$, with weights $w_m$.
Because we are considering linear inequalities, the bound yielded by this strategy for any $\epsilon$ is simply 
\begin{equation}
C_{n}(\epsilon)=\underset{\{w_{m}\}}{{\rm max}}\left[\sum_{m=1}^{n}w_{m} D_n (m) \right],
\label{eq:nondeterministic_bound}
\end{equation}
with the constraints $0 \leq w_m \leq 1$,  $\sum_{m=1}^{n}w_{m}=1$, and $\sum_{m=1}^{n}(\xfrac{m}{n})w_{m}=\epsilon$. By linearity,  the maximum is achieved with at most two nonzero $w_m$s, so the bound $C_n(\epsilon)$ can easily be evaluated numerically for any finite set of observables. 

\begin{figure}
\begin{center}
\includegraphics[width=.9\linewidth]{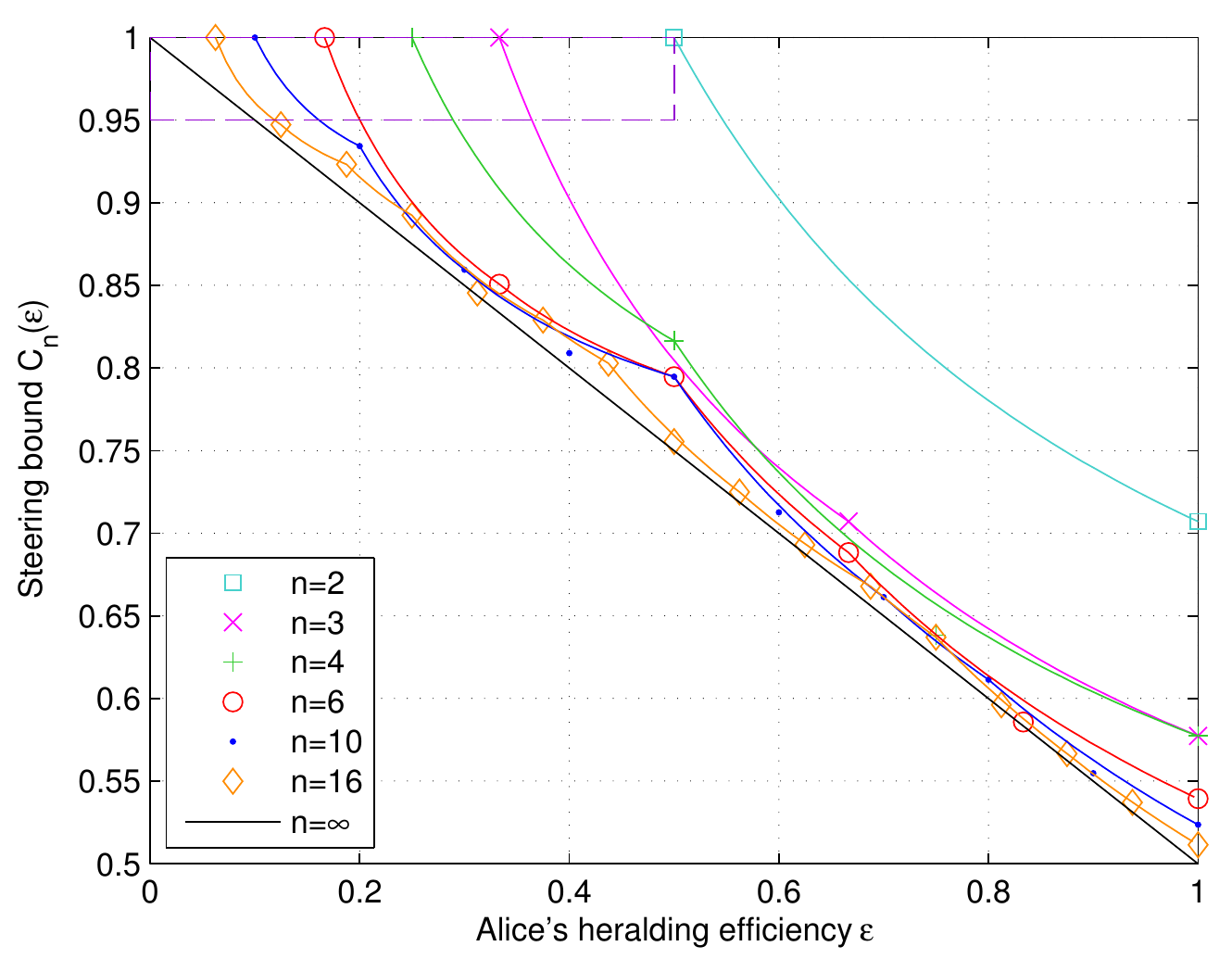}
\end{center}
\caption{ {\bf Loss-dependent EPR-steering bounds.} The solid curves are the theoretical bounds $C_n (\epsilon)$ on $\Str_n$ for demonstrating EPR-steering with no detection loophole, for $n=2, 3, 4, 6, 10, 16,$ and $\infty$. The same-coloured symbols (some of which do not lie on the curves) correspond to the steering parameter $\Str_n$ theoretically obtainable by a cheating Alice using a deterministic strategy (see text).}
\label{fig:Data}
\end{figure}

Note that the optimal cheating ensembles $\{\ket{\xi^{(m)}_i}\}_{p(m)}$ have the same symmetry as the measurement settings (Lemma 1 of Ref.~\cite{Wiseman2007}), which implies that Alice's states are identically arranged around each of Bob's settings. As a consequence, since Alice chooses a state at random from her ensemble, the probability of Alice's claiming a null result is independent of Bob's setting. This is an obvious condition which Bob could place upon Alice's results (to be convinced that Alice's null results really are null results) and for less symmetric setting arrangements (such as the $n=16$ arrangement)  this is an additional condition which could restrict Alice's choice of cheating strategies. Such a restriction can only reduce the effectiveness of Alice's cheating strategy, thereby {\em lowering} the bound $C_n(\epsilon)$ on what she can achieve without entanglement. Thus any demonstration of EPR-steering without such a restriction would remain so with it.

The theoretical values for $C_n(\epsilon)$ are shown in Fig.~\ref{fig:Data}. As expected, $C_n(\epsilon)$ monotonically decreases with $\epsilon$; the $\epsilon=1$ bounds correspond to those derived in Ref.~\cite{Saunders:2010}. The key point is that,  for any arrangement of  $n$ different measurement settings, it is possible to steer using a maximally entangled state if and only if $\epsilon > 1/n$.  This is because $\Str_n$ can reach its maximum value of 1 with maximally entangled states, while the only way for Alice to obtain $\Str_n=1$ by cheating would be to send a state aligned perfectly with one of Bob's measurement directions, and giving a null result for the other $n-1$ settings.

It can finally be shown (see Appendix~\ref{App_C_infty}), that for an infinite number of measurements ($n=\infty$) uniformly distributed on the Bloch sphere, $C_{\infty}(\epsilon) = 1 - \frac {1}{2} \epsilon$. That is, there is a gap between the maximum quantum correlation, $\Str_\infty = 1$, and the EPR-steering bound $C_\infty(\epsilon)$ for any $\epsilon > 0$. Thus, it becomes possible to {demonstrate steering} with {\em  arbitrarily high} losses, as long as Alice and  Bob have a sufficiently high-fidelity singlet state and employ a sufficiently elaborate  many-setting  measurement scheme.

\section{Experimental Demonstration of EPR-Steering}

\subsection{Detection-Loophole-free EPR-Steering}

\begin{figure}
\begin{center}
\includegraphics[width=.9\linewidth]{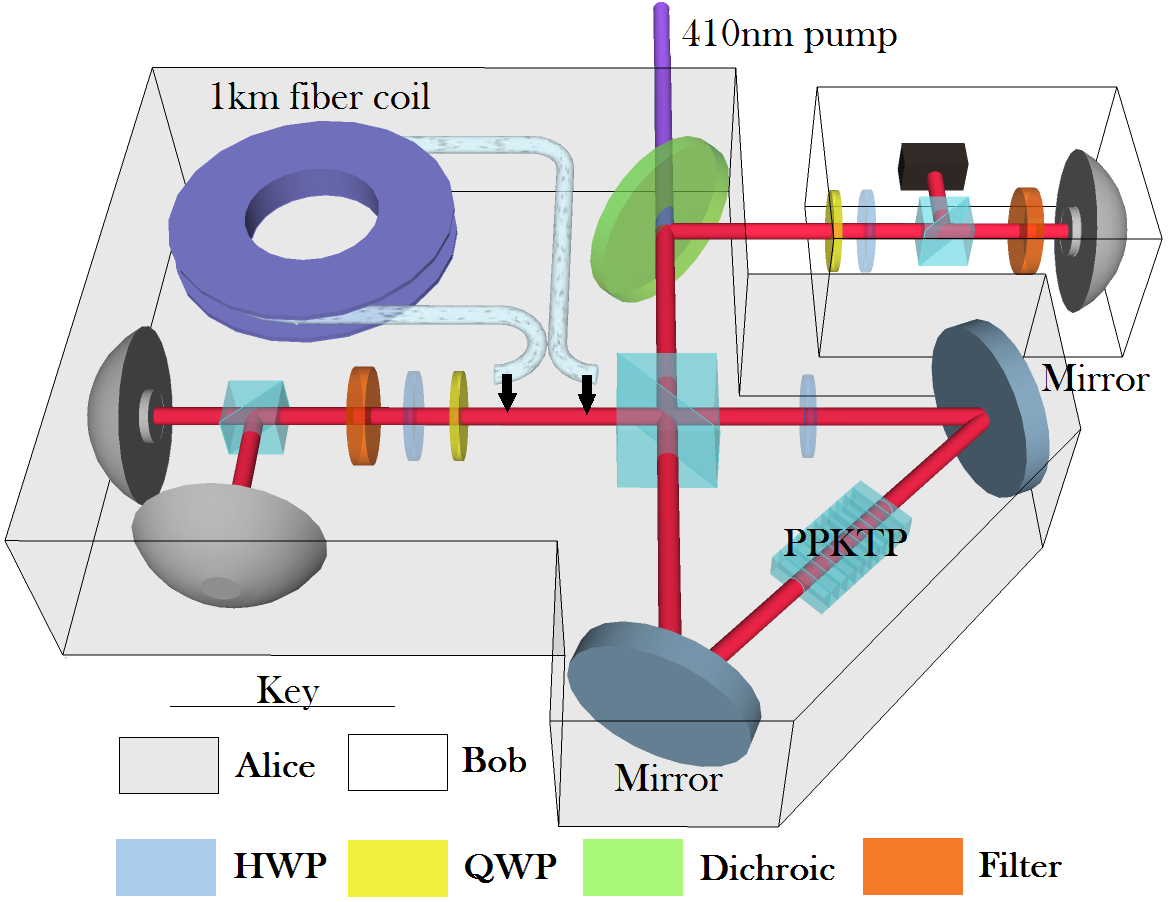}
\end{center}
\caption{{\bf Schematic of the experimental apparatus for demonstrating EPR-steering with no detection loophole.} Bob's apparatus is contained within the white box, while everything else, including the source, is assumed by Bob to be Alice's (grey box) as per Fig.~\ref{fig:protocol}. A 410nm 1mw CW laser pumps a 10mm long periodically-poled KTP (PPKTP) crystal creating the maximally entangled singlet state at 820nm. Measurement settings depend on the orientation of half- and quarter-wave plates (HWP/QWP), mounted in motorised rotation stages, relative to the axes of polarising beam splitters (PBSs), blue cubes. After filtering (Bob: 2nm interference filter, Alice: long  pass filter), photons are coupled to single-mode fibers leading to single-photon-counting modules and counting electronics. For some experiments, we insert a 1~km fiber coil between Alice's detection apparatus and the source. Because Bob trusts his own apparatus, it is sufficient for him to use only one detector (grey hemisphere), corresponding to one (varied at random) of the two eigenstates of his observable $\s{k}^B$. For further details, see Appendix~\ref{App_experimental_details}.}
\label{fig:experimental_setup}
\end{figure}

We experimentally demonstrated detection-loophole free EPR-steering using photonic Bell states generated from an efficient spontaneous parametric down-conversion (SPDC) source---a polarisation Sagnac interferometer, based on Refs.~\cite{Kim2006,Fedrizzi2007} (see  Fig.~\ref{fig:experimental_setup} and Appendix~\ref{App_experimental_details}). Note that in a genuine quantum communication context, Bob must choose his setting independently from one shot to the next. For the purposes of our demonstration this level of rigour was not imposed. In addition,  since we (the experimenters) control Alice's implementation of honest or dishonest strategies, there is no need to force a time ordering of events {\bf 1}--{\bf 3}. In a field deployment, the protocol would require strict time ordering, which could be enforced using an optical delay line for Alice.
 
\begin{figure}
\begin{center}
\includegraphics[width=\linewidth]{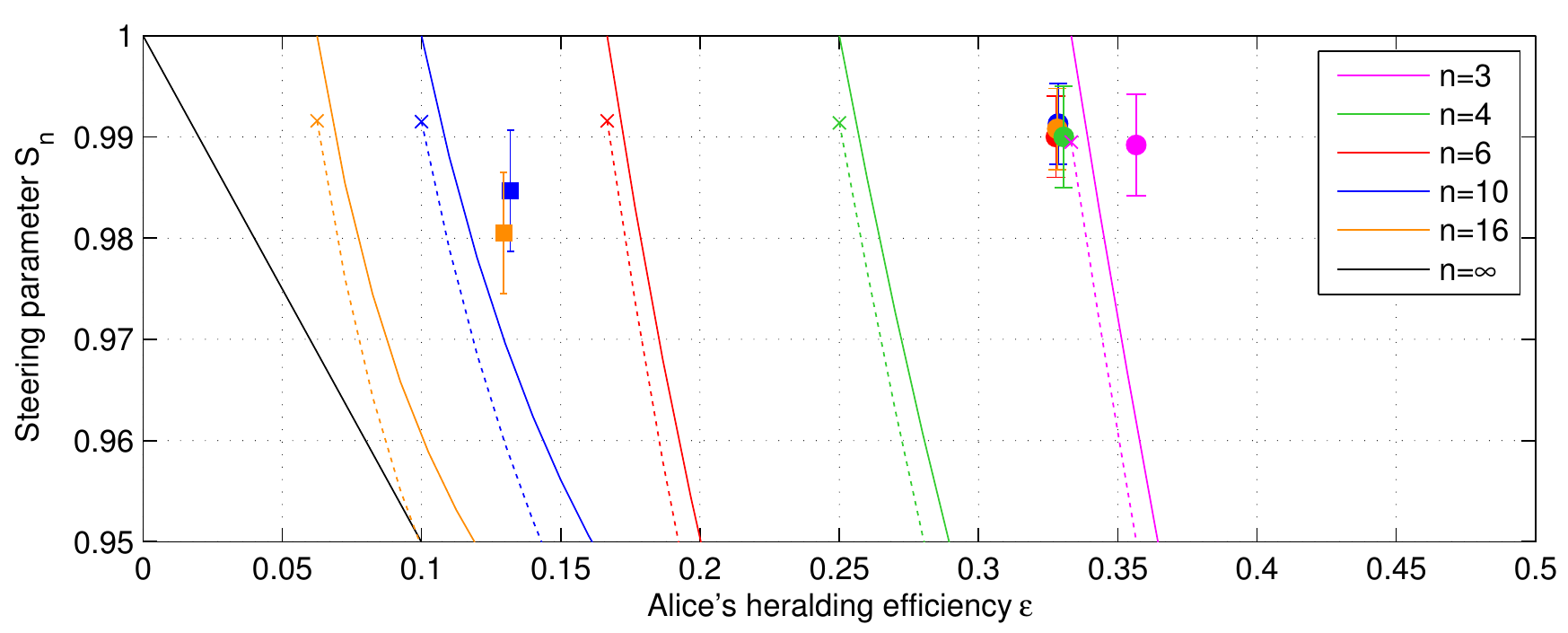}
\end{center}
\caption{{\bf Experimental Demonstration of EPR-Steering.} A zoomed-in section of Fig.~\ref{fig:Data} (dashed purple box) with experimental data included. The data points show the experimental values for the steering parameter  $\Str_n$ obtained  for $n=3, 4, 6, 10$ and $16$ measurement settings. The $\bullet$ points represent data straight from the entangled source, prior to the fiber being installed. The $\blacksquare$ points (for $n=10$ and 16 only) represent data collected after the single mode fiber was installed, demonstrating loss-tolerant EPR-steering with a transmission distance of $1$~km.  The error bars (one standard deviation) take into account systematic measurement errors and Poissonian photon counting noise. The $\times$ points are experimental cheating data (shown in detail in Fig.~\ref{fig:Data_Cheating}), from which we derive our Alice's closest approach using a cheating strategy (dashed curves).}
\label{fig:Data_2}
\end{figure}

A high fidelity maximally entangled state was required to ensure a high value of $\Str_n$. Our tomographically reconstructed state~\cite{James2001} had a fidelity of $0.992\pm0.002$ with the ideal singlet state.  We implemented the $n$-setting measurement schemes for $n= 3, 4, 6, 10$ and $16$, and our experiments yielded values $\Str_n\approx0.99$ for each case (see Fig.~\ref{fig:Data_2}). This gives an absolute ($n=\infty$) lower bound on Alice's required  heralding efficiency of $\epsilon\approx0.02$.  Our source and detector configuration achieved a maximum heralding efficiency of $\epsilon = 0.354\pm0.001$ (as calculated by Bob in the EPR-steering protocol), far above our minimum requirement of $0.02$, enabling us to demonstrate EPR-steering for $n=3$ and greater (Fig.~\ref{fig:Data_2}), with no detection loophole.

\subsection{Experimental Demonstration over 1 km of fiber}

Demonstrating transmission of entanglement over a channel such as an optical fiber is important for real-world  applications such as one-sided DI-QKD. If the entangled source were close to Alice, losses in the line would not be a problem because Bob can post-select on his detected events. But if this were not the case, for instance if  Alice were a mobile field agent and the entangled source were at a base station, line losses to Alice would be critical and loss-tolerant protocols such as ours must be used.

Transmission through a single-mode optical fiber causes the additional problems of  polarisation mode dispersion (PMD)~\cite{Kogelnik2000} and uncompensated birefringence, which reduce $\Str_n$. Thus to test the robustness of our protocol we inserted 1~km of single-mode fiber between the Alice-side output of the Sagnac interferometer and Alice's measurement apparatus (see Fig.~\ref{fig:experimental_setup}). This introduced additional losses of 4.3~dB, and for our source we found $\epsilon = {0.132} \pm 0.001$ for $n=10$ and $\epsilon = {0.130} \pm 0.001$ for $n=16$. We successfully demonstrated  EPR-steering with this setup,  observing $\Str_{10} = 0.985\pm 0.006$ and $\Str_{16} =0.981\pm 0.006$ (Fig.~\ref{fig:Data_2}), 2.6 and 5.3 standard deviations above $C_{10}(\epsilon)$ and $C_{16}(\epsilon)$ respectively.  Based on the intrinsic fiber losses, we estimate that it would still be possible to accomplish the EPR-steering task---with the detection loophole closed, with $n=16$ measurement settings and with all other experimental parameters the same---through $\sim 2$~km of single-mode optical fiber. Thus we can see that an honest Alice can convince Bob that they share entanglement, even in the presence of very significant photon losses.

\subsection{Saturating the cheating bounds using a dishonest Alice}

\begin{figure}
\begin{center}
\includegraphics[width=.9\linewidth]{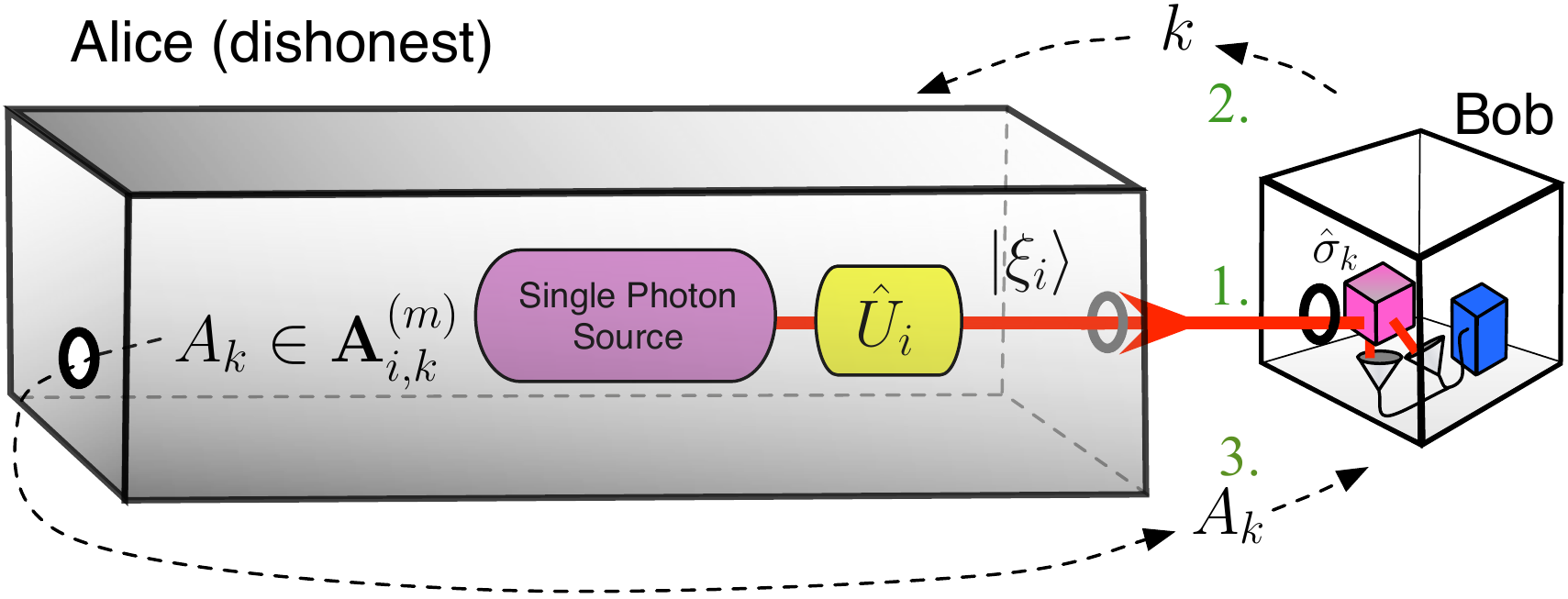}
\end{center}
\vspace{-5ex}
\caption{{\bf Conceptual representation of the EPR-steering task for a dishonest Alice} (to be compared to Fig.~\ref{fig:protocol}). In each round of the protocol, {\bf 1.} Bob receives a photonic qubit, {\bf 2.} announces a measurement setting, $k$, and {\bf 3.} receives a ``measurement'' result from Alice. Bob must assume that Alice controls the source, her line, and her detectors (all enclosed in the grey boxes). In the case of a dishonest Alice, Alice's optimal  ``cheating'' strategy involves sending a single qubit prepared in a pure state $\ket{\xi_i}$ (using a single photon with a polarisation state prepared by the corresponding unitary $\hat{U}_i$), chosen from an optimal set. She announces a ``measurement result'' $A_k$, or a null result (announces nothing), from a look-up table \textbf{A}$^{(m)}_{k,i}$ based on her preparation and Bob's announced measurement direction. Note that the bounds for demonstrating EPR-steering, with no detection loophole, are set precisely to ensure that Alice cannot \textit{actually} cheat --- a dishonest Alice will fail to surpass the upper bound of any EPR-steering inequality.}
\label{fig:protocol_cheating}
\end{figure}

We also approached the EPR-steering protocol experimentally from the point of view of a dishonest Alice, by implementing Alice's optimal cheating strategies, which were determined as described above.

We thus experimentally generated the states in the optimal ``cheating ensembles'', to test for correspondence between $C_n(\epsilon)$ and the maximal correlation $\Str_{n}^\text{cheat}$ attainable by a dishonest Alice. The experimental apparatus for demonstrating Alice's optimal cheating strategy (see Fig.~\ref{fig:protocol_cheating}) involved single qubit state preparation on Alice's side, followed by single qubit measurement on Bob's side. Alice's state preparation involved taking single photons from one arm of a polarisation-unentangled SPDC source. Single-qubit states encoded in polarisation were prepared using a PBS, HWP and QWP, and Bob's measurement device was identical to that used in the case of genuine EPR-steering. The state preparation stage of the cheating experiment lets Alice send any pure state to Bob, while the measurement stage represents Bob's ability to freely draw measurements from the set $\{\hat{\sigma}^B_{k}\}_n$, with $n=3,4,6,10,$ and $16$, as in the case of genuine EPR-steering. Additionally, Bob can implement the $n=2$ settings case, which corresponds to a pair of maximally complementary measurement settings. Alice prepares one of the $p(m)$ states in the optimal deterministic cheating ensemble $\{\ket{\xi^{(m)}_i}\}_{p(m)}$ each of which is (theoretically) equally good at enabling Alice to predict Bob's outcome, given that she is obliged to give a non-null result only for $m$ of Bob's $n$ settings. As explained above, for a given $\epsilon$, Alice's optimal strategy is usually a mixture of two different cheating ensembles ($m'$ and $m''$ say), with weights $w_{m'}$ and $w_{m''} = 1-w_{m'}$. In a noiseless case, Bob's observed $\Str_{n}^\text{cheat}$ is therefore predicted to be  
\bqa
\Str_{n}^\text{cheat} \!\! &=& \!\! \frac{1}{n} \sum_{k=1}^n \, \sum_{m=m',m''} \!\!\! w_m \frac{1}{p(m)} \sum_{i=1}^{p(m)}{\bf A}_{k,i}^{(m)}\bra{ \xi^{(m)}_{i} } \hat{\sigma}^B_{k} \ket{ \xi^{(m)}_{i} } \nonumber \\
& = & C_n(\epsilon).
\eqa
Here ${\bf A}_{k,i}^{(m)}\in \{+1,0,-1\}$ (stored as a look-up table, in which we treat a null result as $0$) is the optimal announcement for Alice given that she has sent state $\ket{\xi^{(m)}_i}$ and Bob has announced that he is measuring along direction $\mathbf{u}_k$.

\begin{figure}
\begin{center}
\includegraphics[width=.9\linewidth]{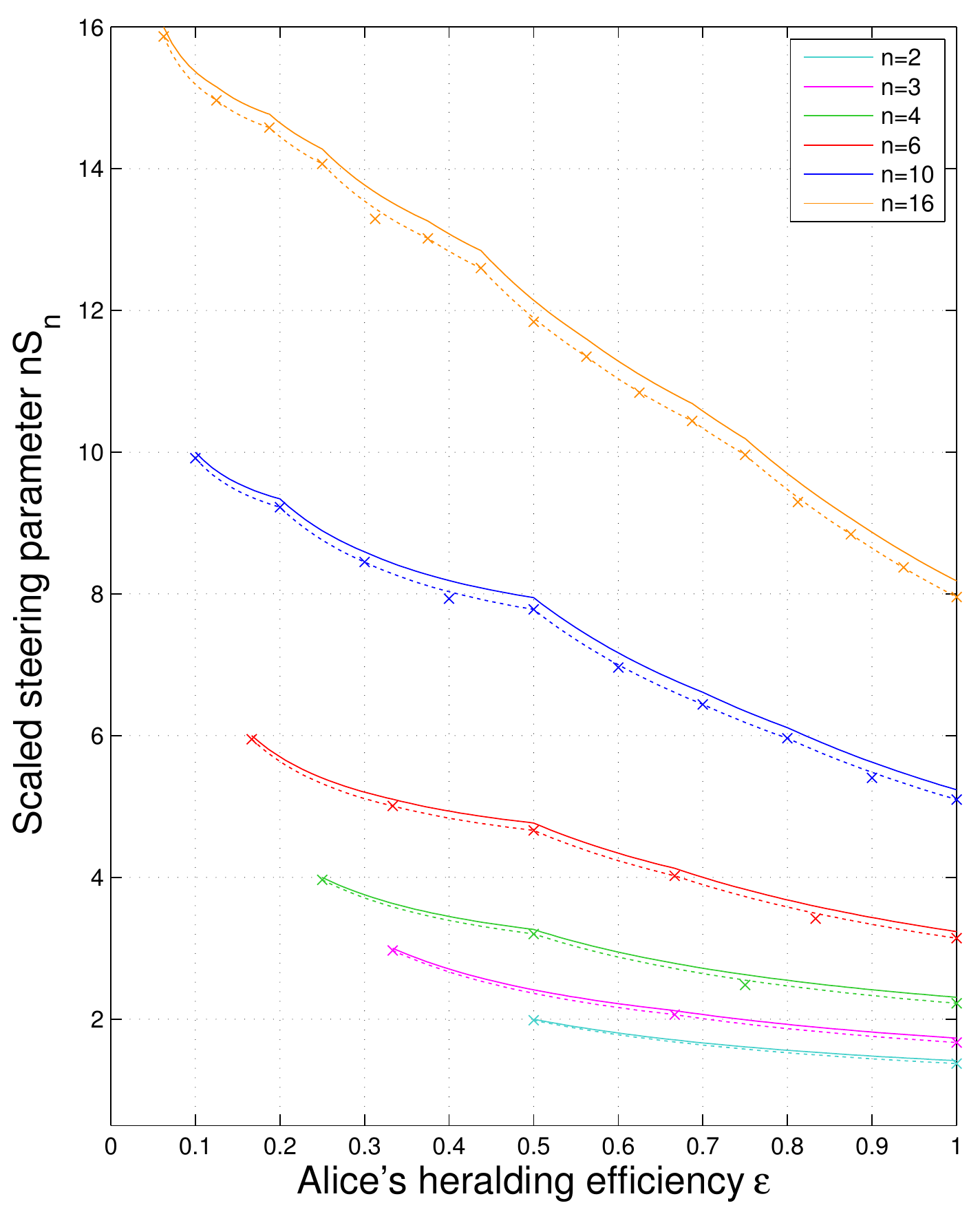}
\end{center}
\caption{{\bf Experimental data for a dishonest Alice.} The solid curves are the bounds $C_n (\epsilon)$ on $\Str_n$, for $n=2, 3, 4, 6, 10$  and $16$. The vertical axis shows a scaled version of the steering parameter, $n \Str_n$, purely for clarity when comparing the different bounds. The same-coloured $\times$'s correspond to the experimentally observed steering parameter $\Str_{n}^\text{cheat}$ obtained by a cheating Alice using a deterministic strategy. The dashed lines, derived from the data points, show the maximum $\Str_{n}^\text{cheat}$ our Alice could achieve by combining two different deterministic strategies to  simulate  a heralding efficiency  $\epsilon$. Error bars are smaller than marker dimensions.}
\label{fig:Data_Cheating}
\end{figure}

Using this technique, we experimentally demonstrated that Alice could indeed come close to saturating (but not exceed) the bounds $C_n(\epsilon)$; see Figs.~\ref{fig:Data_2} and~\ref{fig:Data_Cheating}. The small discrepancies between the measured $\Str_{n}^\text{cheat}$ and the theoretical bound $C_n(\epsilon)$ arose from slightly imperfect state preparation and measurement settings.

\section{Discussion}

We have thus closed the detection loophole in a photonic quantum nonlocality experiment. Our photonic protocol works with arbitrarily large transmission losses---specifically, the novel EPR-steering inequalities we derived allow for arbitrarily low heralding efficiency. We demonstrated the violation of  such  inequalities  over a 1~km optical fiber, with a heralding efficiency for Alice of $-8.9$~dB (13\%). Increasing the number of settings, the state fidelity, or Alice's detection efficiency, would allow for demonstrations of the EPR effect over substantially longer distances.

The ability to keep the EPR-steering detection loophole closed with large losses  opens new possibilities for security in long-range transmission of photonic entanglement over optical fiber, through free space~\cite{Zeilinger07} or to a satellite~\cite{Zeilinger03}. This has potential applications in cryptography, as well as allowing tests of Einstein's ``spooky action'' over unprecedented distances.

\section{Acknowledgments}

We thank Matthew Palsson, Alessandro Fedrizzi and Devin Smith for helpful discussions. This research was conducted by the Australian Research Council Centre of Excellence for Quantum Computation and Communication Technology (Project number CE110001027).

\appendix

\section{Calculation of $C_\infty(\epsilon)$, for infinitely many observables}

\label{App_C_infty}

We consider here the case $n\to \infty$, where Bob use infinitely many observables with uniform distribution on the sphere. 

Let $\theta \in [0,\pi]$ be the angle between Bob's measurement direction and  the pure state   $\ket{\xi}$ sent by a dishonest    Alice, so that the expected correlation between Bob's result and what Alice reports is $\left|\cos(\theta)\right|$. Clearly if Alice is allowed an apparent efficiency $\epsilon$, her optimal strategy is to report a non-null result only when $ \left|\cos(\theta)\right| > \cos \Theta_\epsilon$, where $\Theta_\epsilon $ is the half-angle of a cone which subtends a solid angle $\Omega$ satisfying $\Omega/4\pi = \epsilon/2$. That is, $\cos\Theta_\epsilon = 1-\epsilon$. Using $d\Omega = d(\cos\theta) d\phi$, this optimal strategy gives a correlation, averaged  over the cases where Alice gives a non-null report, of
\bqa
C_\infty(\epsilon) &=& \frac{1}{\epsilon}\int_0^{2\pi} \frac{d\phi}{2\pi} \left\{ \int_{-1}^{-1+\epsilon} + \int_{1-\epsilon}^{1} \right\} 
  \frac{d(\cos \theta)}{2}  \left|\cos(\theta)\right| \nonumber \\
&=& 1 - \frac{\epsilon}{2}, \label{eq:epsilonbound}
\eqa
as mentioned in Section~\ref{sec_calc_bnd}.
Note that this is independent of the state that Alice sends.

\section{Experimental Apparatus}

\label{App_experimental_details}

In this Appendix we provide more technical details on the experimental setup we used to demonstrate detection-loophole-free EPR-steering.

\subsection{Photon Sources}

Our source used a  Toptica iBeam 405 (with 410nm diode) laser, operated with an external diffraction grating (Thorlabs GR25-1204) in the Littrow configuration. The output power after the external grating is 3mW. The grating output is fiber coupled and pumps a 10~mm-long periodically-poled KTP (PPKTP) crystal bidirectionally. The PPKTP crystal is embedded in the Sagnac interferometer\cite{Kim2006,Fedrizzi2007}, giving rise to polarisation-entangled photon pairs at 820~nm via spontaneous parametric down conversion. The Sagnac entangled source can achieve a high heralding efficiency ($\epsilon = 0.354\pm0.001$), because the collinear quasi-phase-matching of the PPKTP crystal provides SPDC modes that are approximately gaussian, so that efficient coupling to single mode fiber is possible. At one output (Alice's side), we use a high-transmission long pass filter to maximize source efficiency, while at the other output (Bob's side) we use a 2~nm interference filter to filter the photons and reject background light. A dichroic mirror separates the down conversion mode from the pump mode in Bob's output arm. The outputs are coupled into single-mode fibers, and connected to Perkin Elmer single photon counting modules (SPCM-AQR-14-FC) and counting electronics. The silicon avalanche photo diodes have a quantum efficiency of approximately 50\% at 820nm. Using a coincidence window of $\sim 3$~ns, a coincidence count rate of approximately 6000 counts per second is achieved. The measured contribution in the coincidence rate from double-pair SPDC emission events is very small, approximately $0.1$ per second.

For the single photon source (used in the experimental implementation of Alice's optimal cheating strategy), one arm of a polarisation-unentangled critically phase matched type-I bismuth triborate downconversion source is used. This was pumped by a 60mW 410nm CW laser. 

\subsection{1~km  transmission channel}

 A 1 km long single-mode fiber at $820$~nm (Thorlabs SM800-5.6-125) was introduced between Alice's output of the Sagnac interferometer and her measurement apparatus. As well as introducing loss, the fiber implements an unknown polarisation unitary operation due to fiber birefingence. We undo this unitary operation in two stages. First we correct the state in the $Z$ basis using a polarisation fiber controller, creating the state $\ket{\psi}=\frac{1}{\sqrt{2}}(\ket{10}+{e^{i \phi}}\ket{01})$. We set the phase, $\phi$, to $\frac{\pi}{2}$ using a tilted half wave plate set at its optic axis. The slight decrease in the steering parameter over the transmission distance is due to a combination of fiber noise (e.g.\ polarisation mode dispersion causing decoherence ~\cite{Kogelnik2000}) and minor errors in performing the polarisation correction.

\section{Experimental Error Calculation}

\label{App_error_calculation}

In order to be sure that we have demonstrated EPR-steering, we need to know that the uncertainty in our measured $\Str_n$ is not so large as to make it possible that the true value would be less than the EPR-steering bound $C_n(\epsilon)$. By ``true value'' we mean the value that would be obtained if all of the assumptions that went into deriving the bound $C_n(\epsilon)$ were satisfied, namely that Bob's measurements are perfect, and that the experiment yields the true quantum averages (which would require an infinite sample size). That is, we need to take into account (1) imperfection of Bob's measurements that could lead to an over-estimation of $\Str_n$ (systematic error), and (2) statistical errors in $\Str_n$. These are determined in Parts 1 and 2 below, respectively. Note that we do not have to worry about systematic errors in Alice's measurement settings, since we make no assumptions about them in order to derive the EPR-steering bound.

\subsection{Experimental error calculation, part 1: systematic error} 

In an ideal experiment, Bob's measurement corresponds to projecting his state onto 
one out of two orthogonal pure states, represented by opposite vectors ${\bf u}_k$ and $-{\bf u}_k$ on the Bloch sphere. Bob's actual measurement will be nonideal in two ways. First, because the manufactured PBS has only a finite extinction ratio, the ``projection operators'' for Bob's measurements will actually comprise a projector mixed with a very small ($\approx 0.01$) amount of the identity operator. This can only ever decrease the correlation with Alice's results, so if Bob takes this effect into account, it can only be to Alice's benefit, by making it easier for her to convince him that she is steering his state. Therefore, to subject our demonstration of steering to the highest level of rigour, we can ignore this imperfection. The second sort of imperfection is that the true states onto which Bob projects, corresponding to vectors ${\bf \tilde u}_k$ and $-{\bf \tilde u}_k'$ on the Bloch sphere, differ slightly from ${\bf u}_k$ and $-{\bf u}_k$ respectively. Note also that in our experiment Bob used only one detector, for reasons of space efficiency; thus he needs to implement two different projections for each choice of setting $k$, and hence ${\bf \tilde u}_k'$ is not necessarily equal to ${\bf \tilde u}_k$. These errors arise from Bob's inability to perform rotations on the Bloch sphere to arbitrary accuracy, for the following reasons: a. due to imperfect alignment of the optic axis of his wave plates (QWP and HWP) with projection axis ($\hat{\sigma}_Z$) of the PBS, b. the repeatability error in the motorised stages (setting the angles of both wave plates), and c. due to wave plate imperfections -- their polarisation retardance is quoted only to within $\pm\pi/250$. The magnitudes of all of these errors is systematically determined, and a Monte Carlo simulation including all of the aforementioned factors allows us to determine the maximum infidelity of Bob's actual measurements (${\bf \tilde u}_k$) with his ideal measurements (${\bf u}_k$). Unlike the error due to a finite PBS extinction ratio, Bob's measurement misalignment can, in principle, make it easier for Alice to fake steering his state. Therefore it is essential to bound the error in our measured $\Str_{n}$ due to this sort of error. 

Because Bob only uses a single detector, we define the outcome $B_k=+ 1$ as being Bob getting a photon and ``discovering'' that he was projecting in the direction $\tilde{\bf u}_k$, and $B_k=- 1$ likewise but projecting in the direction $-\tilde{\bf u}_k'$. 
Provided (as is the case) that Bob chooses to project in the directions $\tilde{\bf u}_k$ and $-\tilde{\bf u}_k'$ with equal probability, if there were no misalignment errors then the rate of occurrence of the event ``$B_k=+1$ or $B_k=-1$'' would be independent of Alice's results.  But in the nonideal situation we cannot make that assumption. Therefore the observed probabilities $\tilde P(A_k,B_k)$ for the four possible coincidences (i.e. postselected on both Alice and Bob detecting a photon) are defined as

\begin{widetext}

\begin{eqnarray}
& \hspace{-.5cm} \tilde P(A_k,B_k\!=\!+1) = R_{{\bf \tilde u}_k}(A_k,B_k\!=\!+1) / {\cal R}_k \,, \quad  \tilde P(A_k,B_k\!=\!-1) = R_{{\bf \tilde u}_k'}(A_k,B_k\!=\!-1) / {\cal R}_k \,, \label{eq:P_of_Ru} \\
& {\mathrm{with}} \quad {\cal R}_k = \sum_{A_k = \pm 1} \big[ R_{{\bf \tilde u}_k}(A_k,B_k\!=\!+1) + R_{{\bf \tilde u}_k'}(A_k,B_k\!=\!-1) \big] \,. \label{Nk_def}
\end{eqnarray}
where $R$ stands for the rate of the corresponding events occurring. 

Let us represent the state Bob receives, conditioned on Alice's output $A_k=\pm 1$, by a vector ${\bf v}_{A_k}^B$ in the Bloch sphere, with $|{\bf v}_{A_k}^B| \leq 1$; note that these states do not depend on Bob's setup, ${\bf \tilde u}_k$ or ${\bf \tilde u}_k'$. The rates $R_{{\bf \tilde u}_k^{(\prime)}}(A_k,B_k)$ are then given by 
\begin{eqnarray}
R_{{\bf \tilde u}_k^{(\prime)}}(A_k,B_k) &=& R_{{\bf \tilde u}_k^{(\prime)}}(A_k) \, P_{{\bf \tilde u}_k^{(\prime)}}(B_k|A_k) \   \propto   \  P_{{\bf \tilde u}_k^{(\prime)}}(A_k) \, \frac{1 + (-1)^{B_k} {\bf \tilde u}_k^{(\prime)} \cdot {\bf v}_{A_k}^B}{2} \,. \label{eq:Ru}
\end{eqnarray}
Note that Alice's marginal probabilities, $P_{{\bf \tilde u}_k^{(\prime)}}(A_k)$, normalized so that they sum to one for $A_k=\pm 1$, do not depend on Bob's measurement setup:  $P_{{\bf \tilde u}_k}(A_k) = P_{{\bf \tilde u}_k'}(A_k) \equiv P(A_k)$;
otherwise Bob could signal instantaneously to Alice. Note also that these may be slightly different from Alice's experimentally observed marginals $\tilde P(A_k)$ calculated from the full postselected distribution  $\tilde P(A_k,B_k)$.

From equation~(\ref{eq:P_of_Ru}) and equation~(\ref{eq:Ru}), one can calculate the experimentally observed correlations $\tilde E_k = \langle A_k \tilde{\sigma}_{k}^{B}\rangle$ (corresponding to the actual measurement ``$\tilde{\sigma}_{k}^{B}$'', rather than the ideal one, $\hat{\sigma}_{k}^{B}$) to be
\begin{eqnarray}
\tilde E_k \ \equiv \ \sum_{A_k,B_k} A_k B_k \,  \tilde P(A_k,B_k) & = & \Big[  P(A_k=+1) \frac{1+ {\bf \tilde u}_k \cdot {\bf v}_{A_k=+1}^B}{2} - P(A_k=-1) \frac{1+ {\bf \tilde u}_k \cdot {\bf v}_{A_k=-1}^B}{2} \nonumber \\[-2mm]
 && \quad -  P(A_k=+1) \frac{1- {\bf \tilde u}_k' \cdot {\bf v}_{A_k=+1}^B}{2} +  P(A_k=-1) \frac{1- {\bf \tilde u}_k' \cdot {\bf v}_{A_k=-1}^B}{2} \Big] / {\cal N}_k \,. \qquad \label{eq:tilde_Ek_decomp}
\end{eqnarray}
Here ${\cal N}_k$ is defined so that the four terms above (without the minus signs) sum to one.
Defining 
\bqa
{\bf \bar u}_k \ \equiv \ \ro{{\bf \tilde u}_k + {\bf \tilde u}_k'}/{2} , &\quad &
{\bf \bar v}_k \ \equiv \ P(A_k=+1) \, {\bf v}_{A_k=+1}^B -  P(A_k=-1) \, {\bf v}_{A_k=-1}^B ,\\
\delta {\bf u}_k \ \equiv \ \ro{{\bf \tilde u}_k - {\bf \tilde u}_k'}/{2} , &\quad &
\delta{\bf v}_k \ \equiv \ P(A_k=+1) \, {\bf v}_{A_k=+1}^B +  P(A_k=-1) \, {\bf v}_{A_k=-1}^B ,
\eqa
we can rewrite $\tilde{E}_k$ more simply as 
\bqa
\tilde E_k =   {\bf \bar u}_k \cdot {\bf \bar v}_k  / {\cal N}_k \qquad {\mathrm{with}} \qquad {\cal N}_k =  1 +  \delta {\bf u}_k \cdot \delta{\bf v}_k \,, \label{eq:Ek}
\eqa
while the ``true value'' $E_k$ of the correlation $\langle A_k \hat{\sigma}_{k}^{B}\rangle$ (corresponding now to the ideal measurement settings $\pm {\bf u}_k$) is simply $E_k = {\bf u}_k \cdot {\bf \bar v}_k$.

In order to quantify the deviation of $\tilde E_k$ from its ``true value'' $E_k$, we characterize the misalignment of the vectors ${\bf \tilde u}_k^{(\prime)}$ by their scalar product with ${\bf u}_k$, the ideal setting: ${\bf \tilde u}_k^{(\prime)} \cdot {\bf u}_k \equiv \chi_k^{(\prime)}$. Further, we assume that we can bound the misalignment by $\chi_k^{(\prime)} \geq X_k > 0$, for some $X_k$ less than, but close to, unity. One can then immediately prove the following, which will be useful later:
\bqa
|{\bf \bar u}_k|^2 + |\delta {\bf \bar u}_k|^2 = 1 \ ; \qquad X_k^2 \ \leq\ |{\bf \bar u}_k|^2 \leq \ 1 \quad {\mathrm{and}} \quad 0 \ \leq\ |\delta {\bf \bar u}_k|^2 \leq \ 1-X_k^2. \label{eq:bounds_u}
\eqa

\medskip

Let us start by bounding the normalisation coefficient ${\cal N}_k$. For that, first note that $|{\bf \bar v}_k|, |\delta{\bf v}_k| \leq 1$, and
\bqa
|{\bf \bar v}_k|^2 + |\delta{\bf v}_k|^2 &=&  2 \,  P(A_k=+1)^2 \, |{\bf v}_{A_k=+1}^B|^2 + 2 \,  P(A_k=-1)^2 \, |{\bf v}_{A_k=-1}^B|^2, \\
& \leq & 2 \,  P(A_k=+1)^2 + 2 \,  P(A_k=-1)^2 \ = \ 1 + (\delta P_k^A)^2 \, , \label{eq:bound_v_v}
\eqa
where  $\delta P_k^A \equiv P(A_k=+1) - P(A_k=-1)$. Defining in a similar way $\delta \tilde P_k^A \equiv \tilde P(A_k=+1) - \tilde P(A_k=-1)$, we find, using equation~(\ref{eq:tilde_Ek_decomp}), ${\cal N}_k \, \delta \tilde P_k^A = \delta P_k^A + \delta {\bf u}_k \cdot {\bf \bar v}_k$. Besides, from equation~(\ref{eq:Ek}) we have $|{\bf \bar u}_k| \, |\bar{\bf v}_k| \geq  {\cal N}_k {\tilde E}_k$.
Hence, following on equation~(\ref{eq:bound_v_v}),
\bqa
|\delta{\bf v}_k|^2 & \leq & 1 + ({\cal N}_k \, \delta \tilde P_k^A - \delta {\bf u}_k \cdot {\bf \bar v}_k)^2 - |{\bf \bar v}_k|^2 \ \leq \ 1 + {\cal N}_k^2 \, (\delta \tilde P_k^A)^2 + 2 \, {\cal N}_k \, |\delta \tilde P_k^A| \ |\delta {\bf u}_k| \, |{\bf \bar v}_k| - |{\bf \bar u}_k|^2 \, |{\bf \bar v}_k|^2 \\
& \leq & 1 + 2 \, {\cal N}_k \, |\delta \tilde P_k^A| \ \sqrt{1-X_k^2} - {\cal N}_k^2 \big[ {\tilde E}_k^2 - (\delta \tilde P_k^A)^2 \big] \,.
\eqa
Now, for typical experimental values, the previous expression decreases with ${\cal N}_k$. From equations~(\ref{eq:Ek}) and~(\ref{eq:bounds_u}), we have ${\cal N}_k \geq 1 - |\delta {\bf u}_k| \, |\delta{\bf v}_k| \geq 1 - \sqrt{1-X_k^2} \, |\delta{\bf v}_k|$, so that we get
\bqa
|\delta{\bf v}_k|^2 & \leq & 1 + 2 \, \Big(1 - \sqrt{1-X_k^2} \, |\delta{\bf v}_k|\Big) \, |\delta \tilde P_k^A| \ \sqrt{1-X_k^2} - \Big(1 - \sqrt{1-X_k^2} \, |\delta{\bf v}_k|\Big)^2 \big[ {\tilde E}_k^2 - (\delta \tilde P_k^A)^2 \big] \\
 & \leq & 1 - {\tilde E}_k^2 + (\delta \tilde P_k^A)^2 + 2 \, |\delta \tilde P_k^A| \, \sqrt{1-X_k^2} \ + 2 \, \sqrt{1-X_k^2} \, {\tilde E}_k^2 \, |\delta{\bf v}_k| \,,
\eqa
where the negative terms we discarded are negligible for our experimental parameters, so do essentially not affect the tightness of the above bound. By resolving the quadratic equation in $|\delta{\bf v}_k|$ above, we further obtain
\beq  |\delta{\bf v}_k| \ \leq \ \sqrt{1-X_k^2} \, {\tilde E}_k^2 + \sqrt{\ro{1-X_k^2} {\tilde E}_k^4 + 1 - {\tilde E}_k^2 + (\delta \tilde P_k^A)^2  + 2 \, |\delta \tilde P_k^A| \, \sqrt{1-X_k^2} } ,
\eeq
which involves only terms obtainable from experimental data. Substituting back into equation~(\ref{eq:Ek}) we obtain 
$|{\cal N}_k - 1| \leq \delta {\cal N}_k$, where 
\beq
\delta {\cal N}_k  \equiv  \left[ \sqrt{1-X_k^2} \, {\tilde E}_k^2 + \sqrt{\ro{1-X_k^2} {\tilde E}_k^4 + 1 - {\tilde E}_k^2 + (\delta \tilde P_k^A)^2  + 2 \, |\delta \tilde P_k^A| \, \sqrt{1-X_k^2} } \right] \sqrt{1-X_k^2} \,. \label{eq:deltaNk}
\eeq

\medskip

Let us now decompose the vectors ${\bf u}_k$ and ${\bf \bar v}_k$ onto ${\bf \bar u}_k$:
\begin{eqnarray}
{\bf u}_k \, = \, \frac{\chi_k + \chi_k'}{2|{\bf \bar u}_k|} \, \frac{{\bf \bar u}_k}{|{\bf \bar u}_k|} + \sqrt{1-\frac{(\chi_k + \chi_k')^2}{4|{\bf \bar u}_k|^2}} \, {\bf \bar u}_k^{\perp,u}, \qquad
{\bf \bar v}_k \, = \, \frac{{\cal N}_k \tilde E_k}{|{\bf \bar u}_k|} \, \frac{{\bf \bar u}_k}{|{\bf \bar u}_k|} + \sqrt{|{\bf \bar v}_k|^2 - \frac{{\cal N}_k^2 \tilde E_k^2}{|{\bf \bar u}_k|^2}} \, {\bf \bar u}_k^{\perp,v}
\end{eqnarray}
where ${\bf \bar u}_k^{\perp,u}$ and ${\bf \bar u}_k^{\perp,v}$ are two unit vectors on the Bloch sphere, both orthogonal to ${\bf \bar u}_k$.
One then gets
\begin{eqnarray}
\Delta E_k & \equiv & | E_k - \tilde E_k | \ = \ | {\bf u}_k \cdot {\bf \bar v}_k - \tilde E_k | \\
&=& \Bigg| \frac{\chi_k + \chi_k'}{2|{\bf \bar u}_k|} \, \frac{{\cal N}_k \tilde E_k}{|{\bf \bar u}_k|} - \tilde E_k + \sqrt{1-\frac{(\chi_k + \chi_k')^2}{4|{\bf \bar u}_k|^2}} \, \sqrt{|{\bf \bar v}_k|^2 - \frac{{\cal N}_k^2 \tilde E_k^2}{|{\bf \bar u}_k|^2}} \, {\bf \bar u}_k^{\perp,u} \cdot {\bf \bar u}_k^{\perp,v} \Bigg| \\  
& \leq & \left| \frac{\chi_k + \chi_k'}{2} \, \frac{{\cal N}_k}{|{\bf \bar u}_k|^2} - 1 \right| |\tilde E_k| + \sqrt{1-X_k^2}  \, \sqrt{1 - {\cal N}_k^2 \tilde E_k^2} \,.
\end{eqnarray}
To bound this further, one can show that $X_k \leq \xfrac{(\chi_k + \chi_k')}{2 |{\bf \bar u}_k|^2} \leq 1/X_k$. Using the bound on ${\cal N}_k$ derived above, we finally obtain
\beq
 \Delta E_k \leq (1 - X_k + \delta {\cal N}_k) |\tilde E_k| / X_k + \sqrt{1 - X_k} \sqrt{1 - (1 - \delta {\cal N}_k)^2 \tilde E_k^2}, 
\eeq
where all of these quantities are experimentally defined. 
Because systematic errors may not be independent, we add them linearly to obtain the total systematic error in $\Str_{n}$ to be at most
\beq
\Delta\Str_n(\text{systematic}) = \frac{1}{n} \sum_{k} \left[  (1 - X_k + \delta {\cal N}_k) |\tilde E_k| / X_k + \sqrt{1 - X_k} \sqrt{1 - (1 - \delta {\cal N}_k)^2 \tilde E_k^2} \right].
\eeq

The size of the different components of this systematic error can be seen in Tables~\ref{Table:syst_error_1} and ~\ref{Table:syst_error_2}.  The indicative  sizes  of the basic experimental parameters used to calculate the systematic error (they vary only slightly with $n$ and $k$) are: $\tilde E_k\approx 0.99$, $1-X_k\approx 2\times 10^{-4}$, and $|\delta \tilde P_k^A|\approx 0.02$, implying $\delta {\cal N}_k\approx 0.002$.

\begin{table}
\caption{Size of error factors contributing to $\Delta\Str_n(\text{systematic})$ --- Fig.~\ref{fig:Data_2} data --- without 1~km  fiber }
\begin{center}
\begin{tabular}{c|c|c|c}
 \hline
 $n$ & $\Delta\Str_n(\text{systematic})$ & $\frac{1}{n}\sum (1 - X_k + \delta {\cal N}_k) |\tilde E_k| / X_k$&$\frac{1}{n}\sum\sqrt{1 - X_k} \sqrt{1 - (1 - \delta {\cal N}_k)^2 \tilde E_k^2}$ \\
\hline
\hline
$n=3$ &0.0049 & 0.00290 & 0.00195  \\
\hline
$n=4$ & 0.0052 & 0.00311 & 0.00206 \\
\hline
$n=6$ & 0.0045 & 0.00272 & 0.00182 \\
\hline
$n=10$ & 0.0045 & 0.00270 & 0.00180 \\
\hline
$n=16$ & 0.0046 & 0.00277 & 0.00185 \\
\hline
\end{tabular}
\end{center}
\label{Table:syst_error_1}
\end{table}

\begin{table}
\caption{Size of error factors contributing to $\Delta\Str_n(\text{systematic})$ --- Fig.~\ref{fig:Data_2} data --- with 1~km fiber}
\begin{center}
\begin{tabular}{c|c|c|c}
 \hline
 $n$ & $\Delta\Str_n(\text{systematic})$ & $\frac{1}{n}\sum (1 - X_k + \delta {\cal N}_k) |\tilde E_k| / X_k$&$\frac{1}{n}\sum\sqrt{1 - X_k} \sqrt{1 - (1 - \delta {\cal N}_k)^2 \tilde E_k^2}$  \\
\hline
\hline
$n=10$ & 0.0057 & 0.00339 & 0.00232 \\
\hline
$n=16$ & 0.0063 & 0.00370 & 0.00256 \\
\hline
\end{tabular}
\end{center}
\label{Table:syst_error_2}
\end{table}

\end{widetext}

\subsection{Experimental error calculation, part 2: statistical error} 

The statistical error component  $\Delta \Str_n (\text{statistical})$ in the total error $\Delta\Str_n$ is a result of having a finite ensemble size. The error in the  total number of Alice--Bob coincident events $N_{c}$ is $\pm \sqrt{N_c}$, as governed by Poissonian statistics. The error $\Delta \Str_n (\text{statistical})$ is determined by simply propagating the error in the counting errors through to the calculation of the joint probabilities $\langle A_k \hat{\sigma}_{k}^{B}\rangle $. This propagation provides the value $\Delta \langle A_k \hat{\sigma}_{k}^{B}\rangle$, and each of these terms contribute in quadrature to $\Delta \Str_n (\text{statistical})$. 

\subsection{Experimental  error calculation, part 3: total error}

Combining the statistical error with the systematic error derived before, we calculate the error in the experimental value of $\Str_{n}$ as
\[
\Delta \Str_n=\sqrt{\Delta\Str_n(\text{systematic})^{2}+\Delta\Str_n(\text{statistical})^2}.
\]

The magnitude of both the systematic ($\Delta\Str_n(\text{systematic})$) and statistical ($\Delta\Str_n(\text{statistical})$) errors in the data presented in Fig.~\ref{fig:Data_2} is shown in Tables~\ref{Table:tot_error_1} and ~\ref{Table:tot_error_2}.

\begin{table}
\caption{Size of systematic and statistical error factors contributing to $\Delta \Str_{n}$ --- Fig.~\ref{fig:Data_2} data --- without 1~km fiber}
\begin{center}
\begin{tabular}{c|c|c|c|c}
 \hline
 $n$ &$\Str_{n}$ &$\Delta \Str_{n}$& $\Delta\Str_n(\text{systematic})$ & $\Delta\Str_n(\text{statistical})$\\
\hline
\hline
$n=3$ & 0.989& 0.0053 & 0.0048 & 0.0022\\
\hline
$n=4$ & 0.990 & 0.0059 & 0.0052 & 0.0029\\
\hline
$n=6$ & 0.990 & 0.0051 & 0.0045 & 0.0023\\
\hline
$n=10$ & 0.991 &0.0049 & 0.0045 & 0.0019\\
\hline
$n=16$ & 0.991 &0.0048 & 0.0046 & 0.0015\\
\hline
\end{tabular}
\end{center}
\label{Table:tot_error_1}
\end{table}

\begin{table}
\caption{Size of systematic and statistical error factors contributing to $\Delta \Str_{n}$ --- Figure~4 data --- with 1~km fiber}
\begin{center}
\begin{tabular}{c|c|c|c|c}
 \hline
 $n$ &$\Str_{n}$ &$\Delta \Str_{n}$& $\Delta\Str_n(\text{systematic})$ & $\Delta\Str_n(\text{statistical})$\\
\hline
\hline
$n=10$ & 0.9847 & 0.0063 & 0.0057 & 0.0028\\
\hline
$n=16$ & 0.9805 & 0.0067 & 0.0063 & 0.0023\\
\hline
\end{tabular}
\end{center}
\label{Table:tot_error_2}
\end{table}


\begin{thebibliography}{99}
\bibitem{note1}  Note that we neither claim nor require that these arrangements give the strongest tests of EPR-steering for a given $n$, and a given heralding efficiency $\epsilon$.
\end{thebibliography}
\end{document}